**Theory of Low Frequency Magnetoelectric Coupling in Magnetostrictive – Piezoelectric Bilayers**

M.I. Bichurin and V.M. Petrov
Department of Engineering Physics, Novgorod State University,
173003 Novgorod, ul. Bolshaya Sankt-Peterburgskaya 41, Russia
G. Srinivasan
Physics Department, Oakland University, Rochester, Michigan 48309-4401

ABSTRACT

A theoretical model is presented for low-frequency magnetoelectric (ME) effects in bilayers of magnetostrictive and piezoelectric phases. A novel approach, the introduction of an interface coupling parameter *k*, is proposed for the consideration of actual boundary conditions at the interface. An averaging method is used to estimate effective material parameters. Expressions for ME voltage coefficients $\alpha'_E = \delta E/\delta H$, where $\delta E$ is the induced electric field for an applied ac magnetic field $\delta H$, are obtained by solving elastostatic and electrostatic equations. We consider both *unclamped* and *rigidly clamped* bilayers and three different field orientations of importance: (i) longitudinal fields ($\alpha'_{E,L}$) in which the poling field *E*, bias field *H* and ac fields $\delta E$ and $\delta H$ are all parallel to each other and perpendicular to the sample plane; (ii) transverse fields ($\alpha'_{E,T}$) for in-plane H and δH parallel to each other and perpendicular to out-of-plane E and δE, and (iii) in-plane longitudinal fields ($\alpha'_{E,IL}$) for all the fields parallel to each other and to the sample plane. The theory predicts a giant ME coupling for bilayers with cobalt ferrite (CFO), nickel ferrite (NFO), or lanthanum strontium manganite (LSMO) for the magnetostrictive phase and barium titanate (BTO) or lead zirconate titanate (PZT) for the piezoelectric phase. Estimates of $\alpha'_E$ are carried out as a function of the interface coupling *k* and volume fraction *v* for the piezoelectric phase. In unclamped samples, $\alpha'_E$ increases with increasing *k*. The strongest coupling occurs for equal volume of the two phases for transverse and longitudinal cases, but a maximum occurs at *v*=0.1 for the in-plane longitudinal case. Upon clamping the bilayer, the ME effect is strengthened for the longitudinal case and is weakened for the transverse case. Other important results of the theory are as follows. (i) The strongest ME coupling is expected for the in-plane longitudinal fields and the weakest coupling for the (out-of-plane) longitudinal case. (ii) In ferrite based composites, $\alpha'_{E,T}$ and $\alpha'_{E,IL}$ are a factor of 2-10 higher than $\alpha_{E,L}$. (iii) The highest ME voltage coefficients are expected for CFO-PZT and the lowest values are for LSMO-PZT. Results of the present model are compared with available data on volume and static magnetic field dependence of $\alpha'_E$. We infer, from the comparison, ideal interface conditions in NFO-PZT and poor interface coupling for CFO-PZT and LSMO-PZT.

PACS Numbers: 75.80.+q; 75.50.Gg; 75.60.-d; 77.65.-j; 77.65.Ly; 77.84.Dy

## I. Introduction

The magnetoelectric (ME) effect is defined as the dielectric polarization of a material in an applied magnetic field or an induced magnetization in an external electric field.[1,2] The induced polarization ***P*** is related to the magnetic field ***H*** by the expression, ***P***=$\alpha$***H***, where $\alpha$ is the second rank ME-susceptibility tensor. A composite of piezomagnetic and piezoelectric phases is expected to be magnetoelectric since $\alpha = \delta P/\delta H$ is the product of the piezomagnetic deformation $\delta z/\delta H$ and the piezoelectric charge generation $\delta Q/\delta z$.[3,4] We are interested in the dynamic ME effect. For an ac magnetic field $\delta H$ applied to a biased and poled sample, one measures the induced voltage $\delta V$. The ME voltage coefficient $\alpha'_E=\delta E/\delta H=\delta V/t\delta H$ and $\alpha = \varepsilon_o \varepsilon_r \alpha'_E$ where *t* is the composite thickness and $\varepsilon_r$ is the relative permittivity.

A similar ME coupling can be accomplished in composites of magnetostrictive and piezoelectric phases.[4-9] Although ferrites are not piezomagnetic, the ac magnetostriction leads to pseudo-piezomagnetic and ME couplings. Composites of interest in the past were bulk samples of $NiFe_2O_4$ (NFO) or $CoFe_2O_4$ (CFO) with $BaTiO_3$ (BTO) and are generally synthesized by sintering the ferrite and $BaTiO_3$ powders.[4] Thin disks of the samples are polarized with an electric field E perpendicularly to its plane. The ME coefficient $\alpha'_E$ are measured for two conditions: (i) transverse fields ($\alpha'_{E,31}$ or $\alpha'_{E,T}$) for *H* and $\delta H$ parallel to each other and to the disk plane (*1,2)* and perpendicular to $\delta E$ (direction-*3*) and (ii) longitudinal fields ($\alpha'_{E,33}$ or $\alpha'_{E,L}$) for all the three fields parallel to each other and perpendicular to sample plane. Sintered bulk composites, in general, show ME coupling much smaller than predicted values.[4,5] The main reason is low resistivity for ferrites that (i) limits the electric field for poling, leading to poor piezoelectric coupling and (ii) generates leakage current through the sample that results in loss of piezoelectrically generated charges.

Problems inherent to bulk samples could be eliminated in a layered structure.[5-11] The strongest ME coupling is expected in a layered structure due to (i) the absence of leakage current and (ii) ease of poling to align the electric dipoles and strengthen the piezoelectric effect. Harshe, et al., proposed such structures, provided a theoretical model for a bilayer and prepared multilayers of CFO-PZT or BTO by sintering thick films.[12] Their samples of CFO-PZT showed weak ME effects and CFO-BTO did not show any ME coupling. We synthesized similar bilayers and multilayers of CFO-PZT and NFO-PZT. The samples were prepared from 10-40 μm thick films obtained by tape-casting. Our studies showed strong ME coupling in CFO-PZT and a giant ME effect in NFO-PZT for transverse fields.[9] Other significant recent developments in this regard concern the observation of ME effects in bilayers and multilayers of lanthanum manganite-PZT and terfenol–PZT with $\alpha'_E$ ranging from 30-4680 mV/cm Oe.[8, 10, 11]

This work focuses on a theoretical understanding of the recent observation of giant ME effects in several bilayers. Previous attempts were concerned with longitudinal ME voltage coefficient $\alpha'_{E,33}$ for the ideal coupling at the interface.[12] The major deficiencies of the model are as follows. (i) For the longitudinal case, influence of the finite magnetic permeability for the ferrite was ignored. A reduction in the internal magnetic field and weakening of ME interactions are expected due to demagnetizing fields. (ii) The model did not consider ME coupling under transverse field orientations for which studies on ferrite-PZT show a giant ME effect. (iii) It is necessary to quantify less-than-ideal interface coupling. Here we present a comprehensive theory in which the composite is considered as a homogeneous medium with piezoelectric and magnetostrictive subsystems.[13-19] Important aspects of the model are as follows.[17] (i) We take into account less-than-ideal interface conditions by introducing an interface coupling parameter $k$. (ii) Expressions for longitudinal and transverse ME coefficients are obtained using the solution of elastostatic and electrostatic equations. (iii) We consider a third field orientation of importance in which $E$, $dE$, $H$, and $dH$ are in the sample plane and parallel to each other and is referred to as in-plane longitudinal ME coupling ($\alpha'_{E,11}$ or $\alpha'_{E,IL}$). (iv) The theory is developed for two types of measurement conditions: unclamped and clamped bilayers. (v) The ME voltage coefficients are estimated from known material parameters (piezoelectric coupling, magnetostriction, elastic constants, etc.,) and are compared with data.

Expressions for $\alpha'_E$ have been obtained for unclamped and clamped bilayers and as a function of interface coupling $k$ and the volume fraction $v$ for the piezoelectric phase. Estimates are for composites consisting of one of the following magnetic oxide: cobalt ferrite due to high magnetostriction, nickel ferrite that shows strong magneto-mechanical coupling, or ferromagnetic lanthanum strontium manganite (LSMO) due to structural homogeneity with PZT. Barium titanate or PZT is considered for the ferroelectric phase. Key findings of the theoretical estimates for unclamped bilayers are as follows. (i) The model predicts a stronger ME coupling in CFO-PZT than in CFO-BTO. (ii) The transverse effect is much stronger than as the longitudinal effect. (iii) $\alpha'_E$ is the highest for in-plane longitudinal field orientations ($\alpha'_{E,IL}$) and is an order of magnitude larger than $\alpha'_{E,L}$. (iv) Estimates of $\alpha'_E$ vs $k$ indicate a decrease in the strength of ME coupling with decreasing $k$. (iv) For $k=1$, $\alpha'_E$ vs $v$ for transverse and longitudinal cases show a maximum for equal volume of ferrite (or manganite) and PZT. But the maximum in $\alpha'_{E,IL}$ vs $v$ occurs for ferrite–rich compositions. (v) When $k$ weakens, the maximum in the ME coupling strength shifts to PZT-rich compositions. (vi) For *rigidly* clamped samples, the theory predicts weakening of the transverse- and strengthening of the longitudinal coupling. The calculated ME susceptibilities are compared with results of earlier models and experimental data. We infer, from the comparison, poor interface coupling in CFO-PZT and LSMO-PZT and excellent coupling in NFO-PZT samples. Detailed theory, application to composites, and comparison with experimental results as provided in the following sections.

## II. General approach

We consider a bilayer in the (*1,2*)-plane consisting of piezoelectric and magnetostrictive phases as shown in Fig.1. Since most studies have dealt with polycrystalline thick films for the two constituents, we do not assume any epitaxial characteristics for the layers. Further, we consider only (symmetric) extensional deformation in this model and ignore any (asymmetric) flexural deformations of the layers that would lead to a position dependent elastic constants and the need for perturbation procedures.[20] An averaging method is used for deriving effective composite parameters and is carried out in two stages.[17-19] In the first stage, the sample is considered as a bilayer. For the polarized piezoelectric phase with the symmetry ∞m, the following equations can be written for the strain and electric displacement:

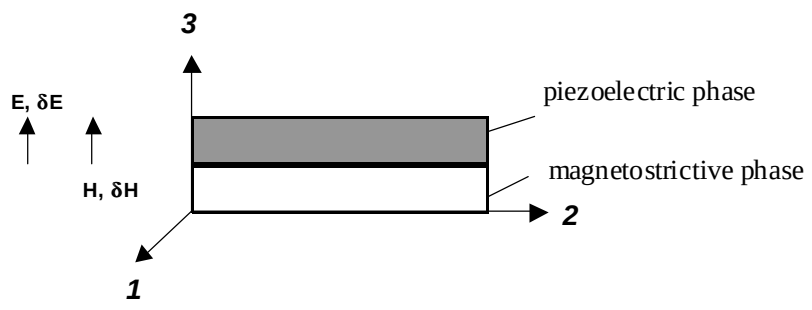

Fig.1: Schematic diagram showing a bilayer of piezoelectric and magnetostrictive phases in the (*1,2*) plane. Field orientations for longitudinal magnetoelectric (ME) effect are also shown.

$$^pS_i = {}^ps_{ij}\,{}^pT_j + {}^pd_{ki}\,{}^pE_k; \qquad (1)$$
$$^pD_k = {}^pd_{ki}\,{}^pT_i + {}^pe_{kn}\,{}^pE_n;$$

where $^pS_i$ and $^pT_j$ are strain and stress tensor components of the piezoelectric phase, $^pE_k$ and $^pD_k$ are the vector components of electric field and electric displacement, ${}^{p}s_{ij}$ and ${}^{p}d_{ki}$ are compliance and piezoelectric coefficients, and ${}^{p}e_{kn}$ is the permittivity matrix. The magnetostrictive phase is assumed to have a cubic symmetry and is described by the equations:

$${}^{m}S_i = {}^{m}s_{ij}\,{}^{m}T_j + {}^{m}q_{ki}\,{}^{m}H_k\,; \qquad (2)$$
$${}^{m}B_k = {}^{m}q_{ki}\,{}^{m}T_i + {}^{m}m_{kn}\,{}^{m}H_n;$$

where ${}^{m}S_i$ and ${}^{m}T_j$ are strain and stress tensor components of the magnetostrictive phase, ${}^{m}H_k$ and ${}^{m}B_k$ are the vector components of magnetic field and magnetic induction, ${}^{m}s_{ij}$ and ${}^{m}q_{ki}$ are compliance and piezomagnetic coefficients, and ${}^{m}m_{kn}$ is the permeability matrix. Equation (2) may be considered in particular as a linearized equation describing the effect of magnetostriction. Assuming in-plane mechanical connectivity between the two phases with appropriate boundary conditions, ME voltage coefficients can be obtained by solving Eqs.(1) and (2). Previous models assumed ideal coupling at the interface.[5] Here we introduce an interface coupling parameter $k =( {}^{p}S_i - {}^{p}S_{i0}) /({}^{m}S_i - {}^{p}S_{i0})$ (i =1,2) where ${}^{p}S_{i0}$ is strain tensor components with no friction between layers. It depends on interface quality and is a measure of differential deformation between piezoelectric and magnetostrictive layers. The coupling factor $k=1$ for an ideal interface is zero for the case with no frictions. The significance of $k$ and its relationship to structural, magnetic and electrical parameters of the composite are discussed later in Section VI.

In the second stage, the bilayer is considered as homogeneous[17-19] and the behavior is described by:

$$S_i = s_{ij}T_j + d_{ki}E_k + q_{ki}H_k;$$
$$D_k = d_{ki}T_i + e_{kn}E_n + a_{kn}H_n; \qquad (3)$$
$$B_k = q_{ki}T_i + a_{kn}E_n + m_{kn}H_n,$$

where $S_i$ and $T_j$ are strain and stress tensor components, $E_k$, $D_k$, $H_k$, and $B_k$ are the vector components of electric field, electric displacement, magnetic field and magnetic induction, $s_{ij}$, $d_{ki}$, and $q_{ki}$ are effective compliance, piezoelectric and piezomagnetic coefficients, and $e_{kn}$, $m_{kn}$ and $a_{kn}$ are effective permittivity, permeability and ME coefficient. Effective parameters of the composite are obtained by solving Eq. (3), taking into account solutions of Eqs. (1) and (2). The mechanical strain and stress for the bilayer and homogeneous material are assumed to be the same and electric and magnetic vectors are determined using open and closed circuit conditions.

## III. Magnetoelectric Coupling in Unclamped Bilayers

### *(i) Longitudinal ME effect*

As shown in Fig.1, we assume (*1,2*) as the film plane and the direction-*3* perpendicular to the sample plane. The bilayer is poled with an electric field $E$ along direction-*3*. The bias field $H$ and the ac field $dH$ are along the same direction as $E$ and the resulting induced electric field $dE$ is estimated across the sample thickness. Then we find an expression for $a'_{E,L}=a'_{E,33}=dE_3/dH_3$. Non-zero components of ${}^{p}s_{ij}$, ${}^{p}d_{ki}$, ${}^{m}s_{ij}$, ${}^{m}q_{ki}$, $s_{ij}$, $d_{ki}$, $q_{ki}$, $a_{kn}$ for this configuration are given in Table 1. Equations 1-3 are then solved for the following boundary conditions:

Table 1. Non-zero coefficients of piezoelectric and magnetostrictive phases and homogeneous material for longitudinal field orientation.

| Piezoelectric phase | |
|---|---|
| Piezoelectric coefficients | Compliance coefficients |
| ${}^{p}d_{15} = {}^{p}d_{24}$<br>${}^{p}d_{31} = {}^{p}d_{32}$<br>${}^{p}d_{33}$ | ${}^{p}s_{11} = {}^{p}s_{22}$<br>${}^{p}s_{12} = {}^{p}s_{21}$<br>${}^{p}s_{13} = {}^{p}s_{23} = {}^{p}s_{31} = {}^{p}s_{32}$<br>${}^{p}s_{33}$<br>${}^{p}s_{44} = {}^{p}s_{55}$<br>${}^{p}s_{66} = 2({}^{p}s_{11} + {}^{p}s_{12})$ |
| **Magnetostrictive phase** | |
| Piezomagnetic coefficients | Compliance coefficients |
| ${}^{m}q_{15} = {}^{m}q_{24}$<br>${}^{m}q_{31} = {}^{m}q_{32}$<br>${}^{m}q_{33}$ | ${}^{m}s_{11} = {}^{m}s_{22} = {}^{m}s_{33}$<br>${}^{m}s_{12} = {}^{m}s_{21} = {}^{m}s_{13} = {}^{m}s_{23} = {}^{m}s_{31} = {}^{m}s_{32}$<br>${}^{m}s_{44} = {}^{m}s_{55} = {}^{m}s_{66}$ |

| Homogeneous material | | |
|---|---|---|
| Piezoelectric coefficients | Piezomagnetic coefficients | Compliance coefficients |
| $d_{15} = d_{24}$<br>$d_{31} = d_{32}$<br>$d_{33}$ | $q_{15} = q_{24}$<br>$q_{31} = q_{32}$<br>$q_{33}$ | ${}^{p}s_{11} = {}^{p}s_{22}$<br>${}^{p}s_{12} = {}^{p}s_{21}$<br>${}^{p}s_{13} = {}^{p}s_{23} = {}^{p}s_{31} = {}^{p}s_{32}$<br>${}^{p}s_{33}$<br>${}^{p}s_{44} = {}^{p}s_{66}$<br>${}^{p}s_{66} = 2({}^{p}s_{11} + {}^{p}s_{12})$ |

$$
\begin{aligned}
&{}^{p}S_i = k\,{}^{m}S_i + (1\text{-}k)\,{}^{p}S_{i0};\quad (i=1,2)\\
&{}^{p}T_i = -\,{}^{m}T_i\,(1\text{-}v)/\,v;\quad (i=1,2)\\
&{}^{p}T_3 = {}^{m}T_3 = T_3;\\
&{}^{m}S_i = S_i;\ (i=1,2)\\
&S_3 = [\,{}^{p}S_3 + {}^{m}S_3(1\text{-}v)]/v;
\end{aligned}
\qquad (4)
$$

where $v={}^{p}v/({}^{p}v+{}^{m}v)$. Here ${}^{p}v$ and ${}^{m}v$ denote the volume of piezoelectric and magnetostrictive phase, respectively, and ${}^{p}S_{10}$ and ${}^{p}S_{20}$ are the strain tensor components at $k=0$.

For finding effective piezoelectric and piezomagnetic coefficients, it is necessary to consider the composite in an electric field $E_3 = V/t$ ($V$ is the applied voltage and $t$ is the thickness of the composite) and a magnetic field $H_3$. The effective $E$ in the piezoelectric and $H$ in the magnetostrictive phases are given by: ${}^{p}E_3 = E_3/v$, ${}^{m}H_3 = (H_3 - v\ {}^{m}B_3/m_0)/(1\text{-}v)$. Using continuity conditions for magnetic and electric fields and open and closed circuit conditions, one obtains the following expressions for effective permittivity, permeability, ME coefficient and longitudinal ME voltage coefficient.

$$
\begin{aligned}
e_{33} = \{&2\,({}^{p}d_{31})^2\,(v\text{-}1) + {}^{p}e_{33}[\,({}^{p}s_{11}+{}^{p}s_{12})(1\text{-}v)+kv({}^{m}s_{11}+\\
&{}^{m}s_{12})]\}\,/\,\{v[({}^{p}s_{11}+{}^{p}s_{12})(1\text{-}v)+\\
&+\,kv({}^{m}s_{11}+{}^{m}s_{12})]\};
\end{aligned}
\qquad (5)
$$

$$
\begin{aligned}
m_{33} = m_0\{&{}^{m}m_{33}[\,kv({}^{m}s_{11}+{}^{m}s_{12}) + (1\text{-}v)({}^{p}s_{11}+{}^{p}s_{12})] -\\
&2kv({}^{m}q_{31})^2\,/\,\{\,m_0\,[v^2({}^{p}s_{11}+{}^{p}s_{12})+\\
&(1\text{--}2v)({}^{p}s_{11}+{}^{p}s_{12})+kv(1\text{-}v)({}^{m}s_{11}+{}^{m}s_{12})]+\\
&{}^{m}m_{33}\{v(1\text{-}v)({}^{p}s_{11}+{}^{p}s_{12})+kv^2({}^{m}s_{11}+{}^{m}s_{12})\text{-}\\
&\text{-}2kv^2({}^{m}q_{31})^2\};
\end{aligned}
\qquad (6)
$$

$$
\alpha_{,33} = \frac{-2m_0kv(1\text{-}v)\,{}^{p}d_{31}\,{}^{m}q_{13}}{[m_0(v\text{-}1)\text{-}{}^{m}m_{33}v][kv({}^{m}s_{11}+{}^{m}s_{12})\text{-}({}^{p}s_{11}+{}^{p}s_{12})(v\text{-}1)]+2\,({}^{m}q_{31})^2\,kv^2}
\qquad (7)
$$

$$
a'_{E,33} = \frac{E_3}{H_3} = 2\frac{m_0 kv(1-v)\,{}^{p}d_{31}\,{}^{m}q_{31}}{\{2\,{}^{p}d_{31}^2(1-v)+{}^{p}e_{33}[({}^{p}s_{11}+{}^{p}s_{12})(v-1)-v({}^{m}s_{11}+{}^{m}s_{12})]\}}\times
\qquad (8)
$$

$$
\times\frac{[({}^{p}s_{11}+{}^{p}s_{12})(v-1)-kv({}^{m}s_{11}+{}^{m}s_{12})]}{\{[m_0(v-1)-{}^{m}m_{33}v][kv({}^{m}s_{12}+{}^{m}s_{11})-({}^{p}s_{11}+{}^{p}s_{12})(v-1)]+2\,{}^{m}q_{31}^2kv^2\}}
$$

Harshe et al [5,12] obtained an expression for longitudinal ME voltage coefficient of the form

$$
a'_{E,33} = \frac{-2v(v\text{-}1)\,{}^{p}d_{13}\,{}^{m}q_{31}}{({}^{m}s_{11}+{}^{m}s_{12})\,{}^{p}e^{T}_{33}\,v + ({}^{p}s_{11}+{}^{p}s_{12})\,{}^{p}e^{T}_{33}\,(1\text{-}v) - 2\,({}^{p}d_{13})^2\,(1\text{-}v)}\ .
\qquad (9)
$$

The above equation corresponds to a special case of our theory in which one assumes ${}^{m}m_{33}/m_0=1$ and $k=1$. Thus the model considered here leads to an expression for the longitudinal ME coupling and allows its estimation as a function of volume of the two phases, composite permeability, and interface coupling. Detailed discussion on the theory and application to specific systems are provided in Section V.

***(ii) Transverse ME effect:***

Table 2. Non-zero coefficients of piezoelectric and magnetostrictive phases and homogeneous material for transverse field orientation

| Piezoelectric phase | | |
|---|---|---|
| Piezoelectric coefficients | Compliance coefficients | |
| ${}^pd_{15} = {}^pd_{24}$<br>${}^pd_{31} = {}^pd_{32}$<br>${}^pd_{33}$ | ${}^ps_{11} = {}^ps_{22}$<br>${}^ps_{12} = {}^ps_{21}$<br>${}^ps_{13} = {}^ps_{23} = {}^ps_{31} = {}^ps_{32}$<br>${}^ps_{33}$<br>${}^ps_{44} = {}^ps_{66}$<br>${}^ps_{66} = 2({}^ps_{11} + {}^ps_{12})$ | |
| Magnetostrictive phase | | |
| Piezomagnetic coefficients | Compliance coefficients | |
| ${}^mq_{35} = {}^mq_{26}$<br>${}^mq_{12} = {}^mq_{13}$<br>${}^mq_{11}$ | ${}^ms_{11} = {}^ms_{22} = {}^ms_{33}$<br>${}^ms_{12} = {}^ms_{21} = {}^ms_{13} = {}^ms_{23} = {}^ms_{31} = {}^ms_{32}$<br>${}^ms_{44} = {}^ms_{55} = {}^ms_{66}$ | |
| Homogeneous material | | |
| Piezoelectric coefficients | Piezomagnetic coefficients | Compliance coefficients |
| $d_{15} = d_{24}$<br>$d_{31} = d_{32}$<br>$d_{33}$ | $q_{35}$ ; $q_{26}$<br>$q_{12}$; $q_{13}$<br>$q_{11}$ | ${}^ps_{11}$ ; ${}^ps_{22}$; ${}^ps_{33}$<br>${}^ps_{12} = {}^ps_{21}$<br>${}^ps_{13} = {}^ps_{31}$; ${}^ps_{23} = {}^ps_{32}$<br>${}^ps_{44}$ ; ${}^ps_{55}$; ${}^ps_{66}$ |

This case corresponds to $E$ and $dE$ along direction-*3* and $H$ and $dH$ along direction-*1* (in the sample plane). Here we estimate the ME coefficient

$$a'_{E,T} = a'_{E,31} = dE_3/dH_1. \quad (10)$$

For this case, non-zero components of ${}^ps_{ij}$, ${}^pd_{ki}$, ${}^ms_{ij}$, ${}^mq_{ki}$, $s_{ij}$, $d_{ki}$, $q_{ki}$, $\alpha_{kn}$ are provided in Table 2. Once again, Eqs.(1)-(3) are solved for boundary conditions in Eq.(4). Expressions for effective permittivity, ME coefficient and transverse ME voltage coefficient are given below.

$$e_{33} = \frac{{}^pd_{31}^2(v-1) + {}^pe_{33}[({}^ps_{11} + {}^ps_{12})(1-v) + kv({}^ms_{11} + {}^ms_{12})]}{v[({}^ps_{11} + {}^ps_{12})(1-v) + kv({}^ms_{11} + {}^ms_{12})]} \quad (11)$$

$$a_{31} = \frac{(v-1)({}^mq_{11} + {}^mq_{21})\,{}^pd_{31}k}{(v-1)({}^ps_{11} + {}^ps_{12}) - kv({}^ms_{11} + {}^ms_{12})}; \quad (12)$$

$$a'_{E,31} = \frac{E_3}{H_1} = \frac{-kv(1-v)({}^mq_{11} + {}^mq_{21})\,{}^pd_{31}}{{}^pe_{33}({}^ms_{12} + {}^ms_{11})kv + {}^pe_{33}({}^ps_{11} + {}^ps_{12})(1-v) - 2k\,{}^pd_{31}^2(1-v)} \quad (13)$$

Equations (13) describe the dependence of ME parameters on volume fraction and is used in Section V to estimate the ME coupling for some representative systems.

***(iii) In-plane longitudinal ME effect:***

Finally, we consider a bilayer poled with an electric field $E$ *in the plane* of the sample. The in-plane fields $H$ and $dH$ are parallel and the induced electric field $dE$ is measured in the same direction (axis-*1*). The ME coefficient is defined as $a'_{E,IL} = a'_{E,11} = dE_1/dH_1$. Expressions for $e$, $m$, $a$ and $a'_E$ obtained from Eqs.(1)-(4) and parameters in Table 3 are given below.

$$e_{11} = {}^me_{11}(1-v) + {}^pe_{11}v + \{k^2v^2(1-v)[{}^pd_{12}^2\,{}^ms_{11} + 2\,{}^pd_{11}^2\,{}^ms_{11}{}^ms_{12}{}^pd_{12}{}^pd_{11} - vk(1-v)^2(2{}^ps_{12}{}^pd_{12}{}^pd_{11} - {}^pd_{11}^2\,{}^ps_{33} - {}^pd_{12}^2\,{}^ps_{11})]\}/[({}^ps_{33}{}^ps_{11} - {}^ps_{12}^2)(1-v)^2 + k^2v^2({}^ms_{11}^2 - {}^ms_{12}^2) + ({}^ps_{33}{}^ms_{11} + {}^ms_{11}{}^ps_{11} - 2{}^ms_{12}{}^ps_{12})kv(1-v)]; \quad (14)$$

$$m_{11} = m_{11}(1-v) + vm_0 + vk(1-v)[(-{}^mq_{11}^2\,{}^ms_{11} - {}^mq_{12}^2\,{}^ms_{11} + 2{}^mq_{11}{}^mq_{12}{}^ms_{12})vk - (1-v)({}^mq_{11}^2\,{}^ps_{33} + {}^mq_{12}^2\,{}^ps_{11} - 2{}^mq_{12}{}^mq_{11}{}^ps_{12})]/[({}^ps_{11}{}^ps_{33} - {}^ps_{23}^2)(1-v)^2 + k^2v^2({}^ms_{11}^2 - {}^ms_{12}^2) + ({}^ps_{33}{}^ms_{11} + {}^ms_{11}{}^ps_{11} - 2{}^ms_{12}{}^ps_{12})kv(1-v)]; \quad (15)$$

$$a_{11}=\{[^mq_{11}(^ps_{33}{}^pd_{11}-{}^ps_{12}{}^pd_{12})+{}^mq_{12}(^ps_{11}{}^pd_{12}-{}^ps_{12}{}^pd_{11})](1-v)+[^mq_{11}(^ms_{11}{}^pd_{11}- {}^ms_{12}{}^pd_{12})+{}^mq_{12}(^ms_{11}{}^pd_{12}-{}^ms_{12}{}^pd_{11})]vk\}vk(1-v)/[(^ps_{33}{}^ps_{11}-{}^ps_{12}{}^2)(1-v)^2+ k^2v^2(^ms_{11}{}^2-{}^ms_{12}{}^2)+(^ps_{33}{}^ms_{11}+{}^ms_{11}{}^ps_{11}-2{}^ms_{12}{}^ps_{13})kv(1-v)]; \quad (16)$$

$$a'_{E,11}=\left(\left({}^mq_{11}\left({}^ps_{33}{}^pd_{11}-{}^ps_{12}{}^pd_{12}\right)+{}^mq_{12}\left({}^ps_{11}{}^pd_{12}-{}^ps_{12}{}^pd_{11}\right)\right)(1-v)+\left({}^mq_{11}\left({}^ms_{11}{}^pd_{11}-{}^ms_{12}{}^pd_{12}\right)+{}^mq_{12}\left({}^ms_{11}{}^pd_{12}-{}^ms_{12}{}^pd_{11}\right)\right)vk\right)vk(1-v)/ /\left(\left((1-p)^me_{11}+v^pe_{11}\right)\left[(1-v)^2\left({}^ps_{11}{}^ps_{33}-{}^ps_{12}^2\right)+(1-v)vk\left({}^ms_{11}{}^ps_{11}+{}^ps_{33}{}^ms_{11}-2{}^ps_{12}{}^ms_{12}\right]+k^2v^2\left({}^ms_{11}^2-{}^ms_{12}^2\right)\right)-vk(1-p)^2[2{}^ps_{12}{}^pd_{11}{}^pd_{12}-{}^ps_{33}{}^pd_{11}^2- {}^ps_{11}{}^pd_{12}^2]+k^2v^2(1-v)\left({}^ms_{11}{}^pd_{12}^2+{}^ms_{11}{}^pd_{11}^2-2{}^ms_{12}{}^pd_{12}{}^pd_{11}\right)\right) \quad (17)$$

The in-plane ME coefficient is expected to be the strongest amongst the cases discussed so far due to high $q$- and $d$- values and the absences of demagnetizing fields. The theory is applied to a series of ferrite-PZT and manganite –PZT in Section V.

Table 3. Non-zero coefficients of piezoelectric and magnetostrictive phases and homogeneous material for in-plane longitudinal field orientation.

| Piezoelectric phase | | |
|---|---|---|
| Piezoelectric coefficients | Compliance coefficients | |
| ${}^pd_{35}={}^pd_{26}$<br>${}^pd_{13}={}^pd_{12}$<br>${}^pd_{11}$ | ${}^ps_{33}={}^ps_{22}$<br>${}^ps_{32}={}^ps_{23}$<br>${}^ps_{13}={}^ps_{23}={}^ps_{31}={}^ps_{32}$<br>${}^ps_{11}$<br>${}^ps_{66}={}^ps_{55}$<br>${}^ps_{44}=2({}^ps_{33}+{}^ps_{32})$ | |
| **Magnetostrictive phase** | | |
| Piezomagnetic coefficients | Compliance coefficients | |
| ${}^mq_{35}={}^mq_{26}$<br>${}^mq_{13}={}^mq_{12}$<br>${}^mq_{11}$ | ${}^ms_{11}={}^ms_{22}={}^ms_{33}$<br>${}^ms_{12}={}^ms_{21}={}^ms_{13}={}^ms_{23}={}^ms_{31}={}^ms_{32}$<br>${}^ms_{44}={}^ms_{55}={}^ms_{66}$ | |
| **Homogeneous material** | | |
| Piezoelectric coefficients | Piezomagnetic coefficients | Compliance coefficients |
| ${}^pd_{35}$; ${}^pd_{26}$<br>${}^pd_{13}$; ${}^pd_{12}$<br>${}^pd_{11}$ | ${}^mq_{35}$; ${}^mq_{26}$<br>${}^mq_{13}$; ${}^mq_{12}$<br>${}^mq_{11}$ | ${}^ps_{11}$; ${}^ps_{22}$; ${}^ps_{33}$<br>${}^ps_{12}={}^ps_{21}$<br>${}^ps_{13}={}^ps_{31}$; ${}^ps_{23}={}^ps_{32}$<br>${}^ps_{44}$; ${}^ps_{55}$; ${}^ps_{66}$ |

## IV. Clamped Bilayers

The theory in Section III is for samples that are free of any external mechanical force. Now we consider bilayers that are subjected to an external force, corresponding to clamped conditions. The force is applied perpendicular to the plane of the bilayer, along direction-$3$. The compliance of clamp system is represented by $s_{c33}$, with zero compliance for rigidly clamped samples and infinite compliance for unclamped samples. Theory for such conditions would facilitate information about the quality of interface coupling between magnetostrictive and piezoelectric layers. As discussed in here and in Section V, clamping leads to significant changes in the strength of ME coupling.

***(i) Longitudinal ME effect***

The boundary conditions for the unclamped case in Sec.III are $T_1 = T_2 = T_3 = 0$. Under clamping, the same conditions become $T_1 = T_2 = 0$ and $S_3 = s_{c33}T_3$, where $s_{c33} = S_3/T_3$ and describes the compliance of clamp system. For unclamped samples $s_{c33} >> s_{33}$, and for rigidly clamped samples $s_{c33} = 0$. The longitudinal ME voltage coefficient is determined by the expression

$$a'_{E,33} = -\frac{a_{33}(s_{33}+s_{c33})-d_{33}q_{33}}{e_{33}(s_{33}+s_{c33})-d_{33}^2}. \quad (18)$$

Various terms in Eq. (18) are given below.

$$s_{33} =[( v(1-v) [2k ({}^{m}s_{12} - {}^{p}s_{13})^{2} - {}^{p}s_{33}( {}^{p}s_{11}+ {}^{p}s_{12}) - k\, {}^{m}s_{11}({}^{m}s_{11} + {}^{m}s_{12})]+ + {}^{m}s_{11} ({}^{p}s_{11} + {}^{p}s_{12}) (2 v -1) - v^{2} k\, {}^{p}s_{33} ({}^{m}s_{11} + {}^{m}s_{12}) - v^{2}\, {}^{m}s_{11}({}^{p}s_{11} + + {}^{p}s_{12})] [( {}^{p}s_{12}+ {}^{p}s_{11}) (v-1) - k v({}^{m}s_{11} + {}^{m}s_{12}) ]; \qquad (19)$$

$$d_{33}= \frac{2\, {}^{p}d_{31}\, k(v-1)({}^{m}s_{12} - {}^{p}s_{13}) + {}^{p}d_{33}[({}^{p}s_{11} + {}^{p}s_{12})(v-1) - k v({}^{m}s_{11} + {}^{m}s_{12})]}{[( {}^{p}s_{12}+ {}^{p}s_{11}) (v-1) - k v({}^{m}s_{11} + {}^{m}s_{12})]^{2}}; \qquad (20)$$

$$q_{33} = \frac{\mu_o (1-v)\{2 k v\, {}^{m}q_{31} ({}^{p}s_{13}-{}^{m}s_{12}) + {}^{m}q_{33}[(1-v)({}^{p}s_{11}+{}^{p}s_{12})+kv({}^{m}s_{11}+{}^{m}s_{12})]\}}{[(1-v)\mu_o+{}^{m}\mu_{33}v] [kv ({}^{m}s_{11} + {}^{m}s_{12}) + (1-v)({}^{p}s_{11} + {}^{p}s_{12}) ] -2 k\, {}^{m}q_{31}^{2} v^{2}} \qquad (21)$$

***(ii) Transverse ME effect***

For clamped sample conditions ($T_1 = T_2 = 0$ and $S_3 = s_{c33}T_3$), the transverse ME voltage coefficient is determined by the following expressions

$$\alpha'_{E,31} = - \frac{\alpha_{31} (s_{33}+s_{c33}) - d_{33}\, q_{13}}{\varepsilon_{33} (s_{33}+s_{c33}) - d_{33}^{2}} \qquad (22)$$

where $s_{33}$ and $d_{33}$ are defined by Eqs. (19) and (20), and

$$q_{13} = - \{v^{2}\, {}^{m}q_{12}[({}^{p}s_{11} + {}^{p}s_{12}) - k({}^{p}s_{13} + {}^{m}s_{11})] + {}^{m}q_{12} (1 - 2v)({}^{p}s_{11} + {}^{p}s_{12}) + + k v\, {}^{m}q_{12}( {}^{p}s_{13} + {}^{m}s_{11}) + k v(1-v)\, {}^{m}q_{11}({}^{p}s_{13} - {}^{m}s_{12})\} [( {}^{p}s_{12}+ {}^{p}s_{11}) (v-1) - k v({}^{m}s_{11} + {}^{m}s_{12}) ]. \qquad (23)$$

***(iii) In-plane longitudinal ME effect***

For clamped samples ($T_1 = T_2 = 0$ and $S_3 = s_{c33}T_3$), one finds the ME voltage coefficient of the form

$$\alpha'_{E,11} = - \frac{\alpha_{11} (s_{33}+s_{c33}) - d_{13}\, q_{13}}{\varepsilon_{33} (s_{33}+s_{c33}) - d_{13}^{2}} \qquad (24)$$

where $s_{33}$ and $q_{13}$ are defined by Eqs. (19) and (23) and

$$d_{13}=\{{}^{p}d_{13}\, v\{k\, {}^{p}s_{13}[ {}^{p}s_{11}(v-1)^{2}+vk\, {}^{m}s_{11}(1-v)]+ {}^{m}s_{12}k[kv[ {}^{m}s_{12}-( {}^{p}s_{13}+ {}^{m}s_{11})(1-v)]- {}^{p}s_{11}(1-v)^{2}+ {}^{p}s_{13}(1-v^{2}+ +kv^{2})]- {}^{p}s_{11}\, {}^{p}s_{33} (v-1)^{2} - {}^{p}s_{13}^{2}(v-1)^{2}(k-1) +k v(v-1)\, {}^{m}s_{11}( {}^{p}s_{11}+ {}^{p}s_{33}) - k^{2}v^{2}\, {}^{m}s_{11}^{2}\}+ {}^{p}d_{11}\, vk\{ {}^{m}s_{12}\, v[({}^{m}s_{12} - - {}^{p}s_{13}- {}^{m}s_{11}) k(1-v)- {}^{p}s_{33} (v-2+k)]-(-1+v)\, {}^{p}s_{13} [k v\, {}^{m}s_{11}+({}^{m}s_{12}+ {}^{p}s_{33}- {}^{p}s_{13}) (1-v)]\}\} / [(1-v)^{2} ({}^{p}s_{13}^{2} - - {}^{p}s_{11}\, {}^{p}s_{33}) + k^{2} v^{2} ({}^{m}s_{12}^{2} - {}^{m}s_{11}^{2}) -v(1-v) k({}^{m}s_{11} ({}^{p}s_{11} + {}^{p}s_{33}) - 2\, {}^{p}s_{13}\, {}^{m}s_{12})]. \qquad (25)$$

It is intuitive to consider two limiting cases, unclamped and rigidly clamped samples. For the first case, corresponding to $s_{c33} >> s_{33}$, Eqs.(18)-(24) reduce to Eqs.(8)-(17). For rigidly clamped samples, i.e., $s_{c33} = 0$, we obtain the following expressions.

$$\alpha'_{E,33} = - \frac{\alpha_{33}\, s_{33} - d_{33}\, q_{33}}{\varepsilon_{33}\, s_{33} - d_{33}^{2}} \qquad (26)$$

$$\alpha'_{E,31} = - \frac{\alpha_{31}\, s_{33} - d_{33}\, q_{13}}{\varepsilon_{33}\, s_{33} - d_{33}^{2}} \qquad (27)$$

$$\alpha'_{E,11} = - \frac{\alpha_{11}\, s_{33} - d_{13}\, q_{13}}{\varepsilon_{11}\, s_{33} - d_{13}^{2}} \qquad (28)$$

Equations (26)-(28) are applied to a series of ferrite-PZT bilayers in Section V.

## V. Application: ME effects in representative bilayers

The preceding comprehensive theoretical treatment resulted in expressions for ME voltage coefficients for three field orientations of importance: longitudinal, transverse, and in-plane longitudinal fields. The most significant features of the current model are as follows. (i) The consideration of three different field configurations for two different sample conditions, unclamped and clamped, so that data on ME coupling could be used for the determination of a unique, single valued interface parameter *k*. Thus the model facilitates quantitative characterization of the interface. (ii) Consideration of a new field configuration, i.e., in-plane longitudinal fields that is shown (in this section) to give rise to a very strong ME coupling. (iii) The model takes into account finite magnetic permeability for the magnetostrictive phase which was ignored in previous studies.[5,12]

Table 4

Material parameters (compliance coefficient s, piezomagnetic coupling q, piezoelectric coefficient d, permeability μ, and petmittivity ε) for lead zirconate titanate (PZT), cobalt ferrite (CFO), barium titanate (BTO), nickel ferrite (NFO), and lanthanum strontium manganite used for theoretical values in Fig.1-7. [Refs.12, 22]

| Material | $s_{11}$ ($10^{-12}$ $m^2/N$) | $s_{12}$ ($10^{-12}$ $m^2/N$) | $s_{13}$ ($10^{-12}$ $m^2/N$) | $s_{33}$ ($10^{-12}$ $m^2/N$) | $q_{33}$ ($10^{-12}$ m/A) | $q_{31}$ ($10^{-12}$ m/A) | $d_{31}$ ($10^{-12}$ m/V) | $d_{33}$ ($10^{-12}$ m/V) | $\mu_{33}/\mu_0$ | $\varepsilon_{33}/\varepsilon_0$ |
|---|---|---|---|---|---|---|---|---|---|---|
| PZT | 15.3 | -5 | -7.22 | 17.3 | - | - | -175 | 400 | 1 | 1750 |
| BTO | 7.3 | -3.2 | | | | | -78 | | 1 | 1345 |
| CFO | 6.5 | -2.4 | | | -1880 | 556 | - | - | 2 | 10 |
| NFO | 6.5 | -2.4 | | | -680 | 125 | - | - | 3 | 10 |
| LSMO | 15 | -5 | | | 250 | -120 | - | - | 3 | 10 |

Now we apply the theory for the calculation of ME coupling in representative layered systems. First we discuss a system of primary interest in the past: cobalt ferrite and barium titanate. Since the strength of $a¢_{E}$ depends sensitively on the concentration of the two phases, we estimated the longitudinal and transverse voltage coefficients as a function of the volume fraction *v* for the piezoelectric phase in CFO-BTO. Results in Fig.2 were obtained assuming ideal interface coupling (*k*=1) and for material parameters given in Table 4.[5,12,21,22] Consider the *v* dependence of $a¢_{E,33}$ in Fig.2. The ME coupling is absent in pure CFO or BTO. As *v* is increased from 0, $a¢_{E,33}$ increases to a maximum for $v_m$=0.4. The coupling weakens with further increase in *v*. Values of $a¢_{E,33}$ based on the model by Harshe[5,12] are also shown for comparison and are 30-40% higher than present values due to implied assumption that $m_{33}=m_0$. It is thus obvious from Fig.2 that demagnetizing fields associated with longitudinal orientation lead to a reduction in $a¢_{E,33}$.

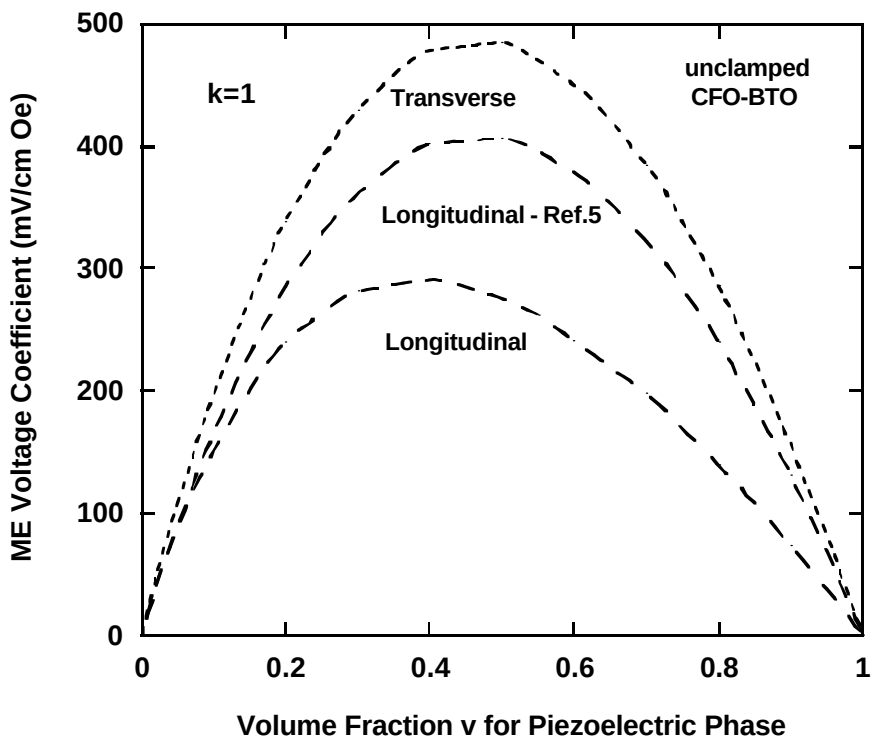

Fig.2: Transverse magnetoelectric (ME) voltage coefficient $a¢_{E,31}=dE_3/dH_1$ and longitudinal coefficient $a¢_{E,33}=dE_3/dH_3$ for a perfectly bonded (*k*=1) two-layer structure consisting of cobalt ferrite (CFO) and barium titanate (BTO). The poling field E is along the sample thickness (direction-*3*). For the transverse case, the bias field H and the ac magnetic field $dH_1$ are assumed to be parallel to each other and to the sample plane and the induced electric field $dE_3$ is measured perpendicular to the sample plane. For the longitudinal case, all the fields are along the direction-3. Estimates are based on the present work and earlier model in Ref.5. Material parameters used for the calculation are given in Table 4.

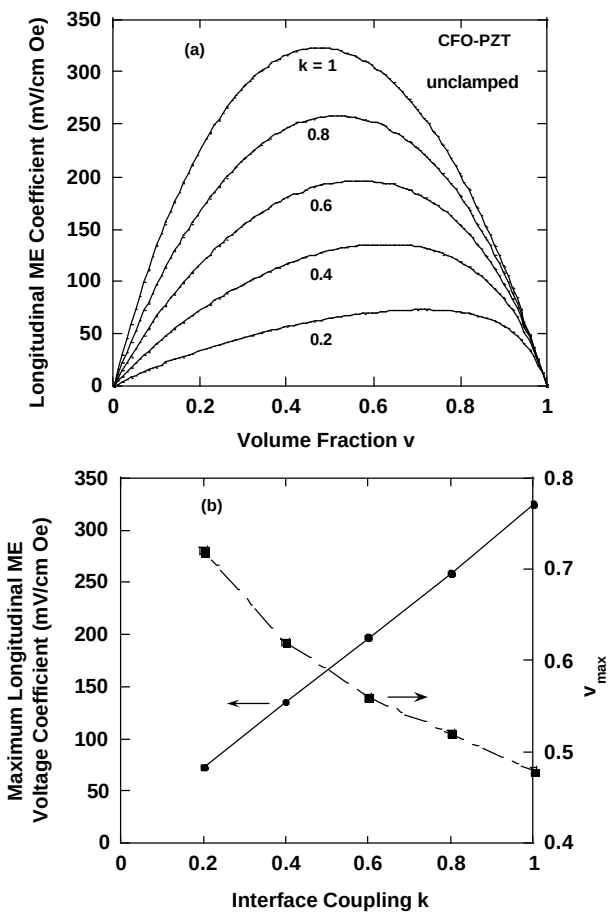

Fig. 3: (a) Estimated dependence of longitudinal ME voltage coefficient on interface coupling $k$ and volume fraction $v$ for CFO-lead zirconate titanate (PZT) bilayer. (b) Variation with $k$ of maximum $a¢_{E,33}$ and the corresponding $v_{max}$.

For transverse fields, one observes similar features in Fig.2, but the maximum in $a¢_{E,31}$ is almost a factor of two higher than $a¢_{E, 33}$. The transverse coefficient peaks at a slightly higher $v$ compared to the longitudinal case. Thus the key are (i) the prediction of a giant ME effect in CFO-BTO, (ii) a transverse coupling much stronger than the longitudinal case, and (iii) maximum coupling for approximately equal volume for the two phases. Now we compare these results with theoretical estimates for bulk composites. Such estimates for bulk composites are based on shape and distribution (connectivity) of the ferrite and barium titanate particles.[12] For $v$=0.5, $a¢_{E, 33}$ ranges from 800 to 1200 mV/cm Oe (such high $a¢_{E,33}$ have never been accomplished in any bulk samples because of leakage current due to low resistivity for ferrites). The current model predicts $a¢_{E,33}$ values comparable to bulk composites of CFO-BTO. The estimated $a¢_{E}$ are orders of magnitude higher than reported values in single phase ME materials.[23]

The bilayer of importance, however, is CFO-PZT due to high piezoelectric coefficients for PZT. We calculated the longitudinal ME coupling in the system and its dependence on the interface coupling parameter $k$ for parameters in Table 4. The variation of $a¢_{E,33}$ with $v$ for a series of $k$-values is shown in Fig.3(a). The strength of $a¢_{E,33}$ decreases as the coupling k weakens and $v_{\max}$ shifts to PZT-rich compositions. Figure 3(b) shows the $k$-dependence of peak values of $a¢_{E,33}$ and the corresponding $v_{\max}$. With increasing $k$ (i) a near-linear increase occurs in the maximum value for $a¢_{E,33}$ and (ii) the ME coupling reaches maximum at progressively decreasing $v_{max}$. As expected, $a¢_{E,33}$ is higher than in CFO-BTO and is due to strong piezoelectric coupling $d_{31}$ for PZT.

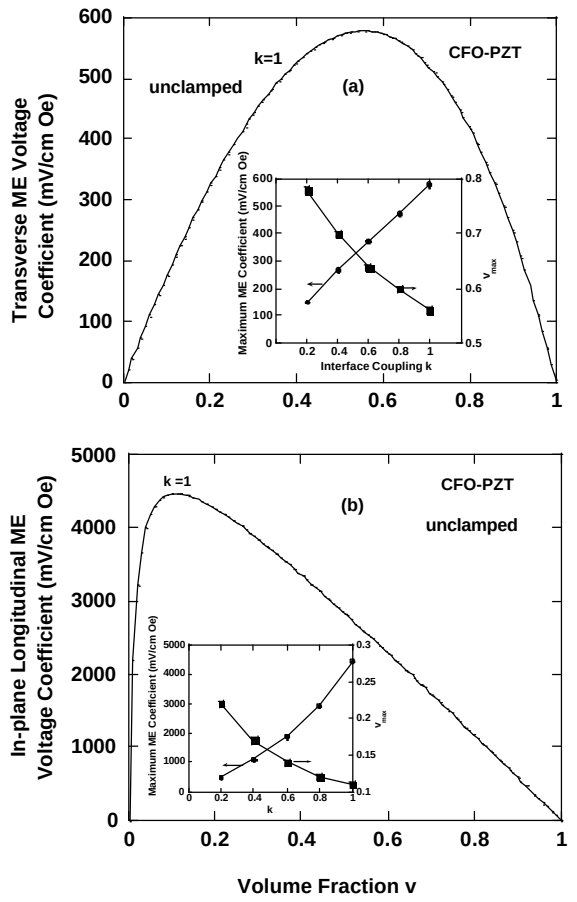

Fig. 4: Results as in Fig.3 for CFO-PZT bilayers but for (a) transverse fields ($a¢_{E,31}$) and (b) in-plane longitudinal fields ($a¢_{E,11}$). For in-plane longitudinal fields, the poling field and other dc and ac fields are parallel to each other and in the sample plane.

Now we consider the ME effect in CFO-PZT for the two other field orientations, transverse and in-plane longitudinal cases. The $v$-dependence of $a¢_{E}$ for both cases is shown in Fig.4 for ideal interface coupling ($k$=1). The insets show variations in $a¢_{E,max}$ and $v_{max}$ with $k$. For transverse fields, peak $a¢_{E}$ is higher by 40% compared to $a¢_{E,33}$. One could attribute this to a strong parallel piezomagnetic coupling $q_{11}$ compared to $q_{31}$ (that determines $a¢_{E,33}$). The transverse ME coupling shows a higher $v_{max}$ compared to the longitudinal orientation. Other features, including those in $a¢_{E,\max}$ vs $k$ and $v_{max}$ vs $k$, are similar in nature to the longitudinal fields. The most significant outcome of the current model is the prediction of the strongest coupling for in-plane longitudinal fields as shown in Fig.4(b). When the fields are switched from longitudinal to in-plane longitudinal orientation, there is a dramatic increase in $a¢_{E}$. The $v$-dependence of $a¢_{E,11}$ shows a rapid increase in ME coupling strength to a maximum for $v$=0.11 and is followed by a near-linear decrease for higher $v$. The voltage coefficient $a¢_{E,max}$ increases from 325 mV/cm for longitudinal orientation to a giant value of 4500 mV/cm Oe. Such an enhancement is understandable due to (i) the absence of demagnetising fields and (ii) increased piezoelectric and piezomagnetic coupling coefficients compared to longitudinal fields (Eq.17). The down-shift in $v_{max}$, from 0.5-0.6 for longitudinal and transverse fields to a much smaller value of 0.1, is related to the concentration dependence of the effective permittivity (Eq. 14).

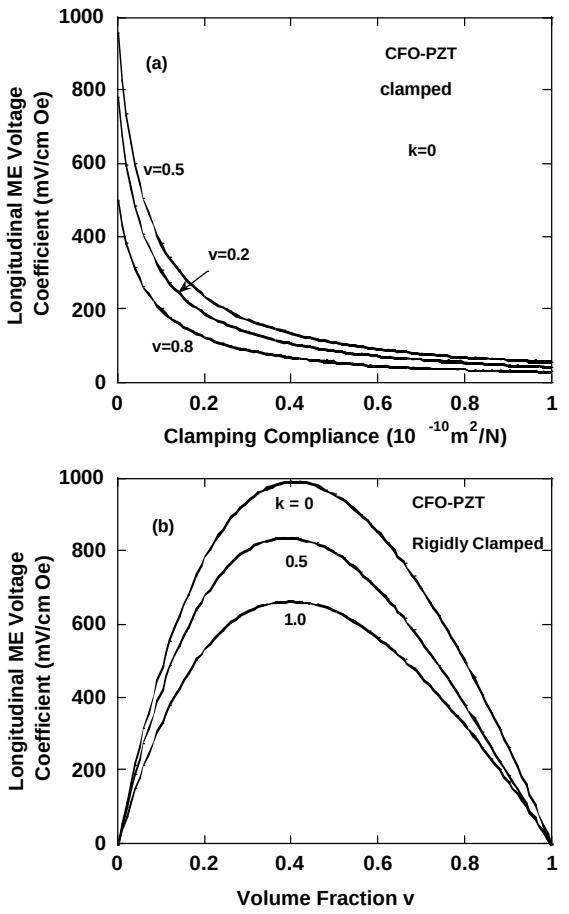

Fig. 5: (a) Variation of the longitudinal ME voltage coefficient for CFO-PZT with the compliance coefficient $s_{c33}$ for the clamping system. An infinite compliance of clamping corresponds to unclamped samples and zero compliance represents a rigidly clamped bilayer. Values are for $k$=0 and for a series of volume fraction $v$ for PZT. (b) $\alpha'_{E,33}$ vs $v$ as a function of $k$ for a rigidly clamped CFO-PZT.

Consider now the effect of clamping a bilayer of CFO-PZT. Very significant changes in the nature of ME coupling are expected when the bilayer is subjected to a uniform stress perpendicular to the sample plane. Representative results for longitudinal fields are shown in Fig.5. Variations in $\alpha'_{E,33}$ with the compliance parameter $s_{c33}$ is depicted in Fig.5(a) for a series of $v$-values. The results are for interface coupling $k$=0. One expects $\alpha'_{E,33}$=0 for $k$=0 in unclamped ($s_{c33}$=∞) bilayers. As the uniaxial stress on the sample is increased ($s_{c33}$ is decreased), $\alpha'_{E,33}$ increases and reaches a peak value for rigidly clamped samples. The enhancement is rather large in samples with equal volume for the two phases ($v$=0.5). Figure 5(b) shows estimated $\alpha'_{E,33}$ vs $v$ for rigidly clamped CFO-PZT for a series of $k$-values. It is quite intriguing that clamping associated enhancement in ME coupling is very high in samples with weaker interface coupling $k$. Key inferences from Fig.3 and 5 are: (i) clamping in general leads to an increase in $\alpha'_{E,33}$ and (ii) the largest increase occurs for rigidly clamped samples with the weakest interface coupling k.

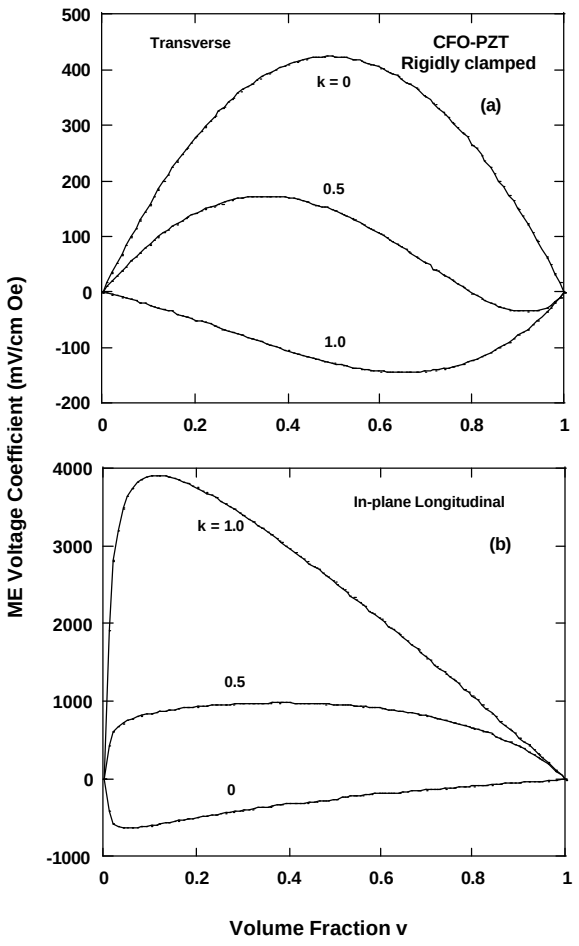

Fig. 6: Dependence of transverse and in-plane longitudinal ME voltage coefficients on volume fraction $v$ and interface coupling $k$ for rigidly clamped (compliance $s_{c33}$ = 0) CFO-PZT bilayer. Negative values for $\alpha'_E$ represent a 180 deg. phase difference between δE and δH.

Similar results on clamping related effects on $\alpha'_E$ for transverse and in-plane longitudinal fields are shown in Fig.6. Estimates of $\alpha'_E$ as a function of $v$ are for representative $k$-values and for rigidly clamped ($s_{c33}$=0) bilayers. For transverse fields, one observes in Fig.6(a) a substantial reduction in $\alpha'_{E,31}$ compared to unclamped samples (Fig.4). The negative value of $\alpha'_{E,31}$ for $k$=1 is just an indicator of the phase difference between $dE_3$ and $dH_1$. A phase reversal is also evident in $\alpha'_{E,31}$ vs $v$ for $k$=0.5. One expects the highest $\alpha'_E$ for $k$=0, as for the longitudinal case. The overall effect of clamping is a reduction in the strength of transverse ME voltage coefficient. Finally, results in Fig. 6(b) for in-plane longitudinal fields show clamping related changes in ME coupling strength that are quite weak compared to longitudinal or transverse cases.

Another bilayer of importance is nickel ferrite-PZT. Although NFO is a soft ferrite with a much smaller anisotropy and magnetostriction than for CFO, efficient magneto-mechanical coupling in NFO-PZT gives rise to ME voltage coefficients comparable CFO-PZT. Using the current model, we estimated $\alpha'_E$ for NFO-PZT for various field orientations and conditions as for CFO-PZT. Representative results for unclamped and rigidly clamped samples with ideal interface coupling ($k$=1) are presented in Fig.7. The most important inferences from the results are as follows. (i) For unclamped bilayers, $\alpha'_E$ is the smallest for longitudinal fields and is the highest for in-plane longitudinal fields. (ii) $\alpha'_{E,31}$ and $\alpha'_{E,11}$ are higher than $\alpha'_{E,33}$ by a factor of 5-30. (iii) Upon rigidly clamping the bilayer, there is a five-fold increase in $\alpha'_{E,33}$, a 50% reduction in $\alpha'_{E,31}$ and a very small decrease in $\alpha'_{E,11}$. Other features in Fig.7 are qualitatively similar to results in Figs.2-6 for CFO-PZT.

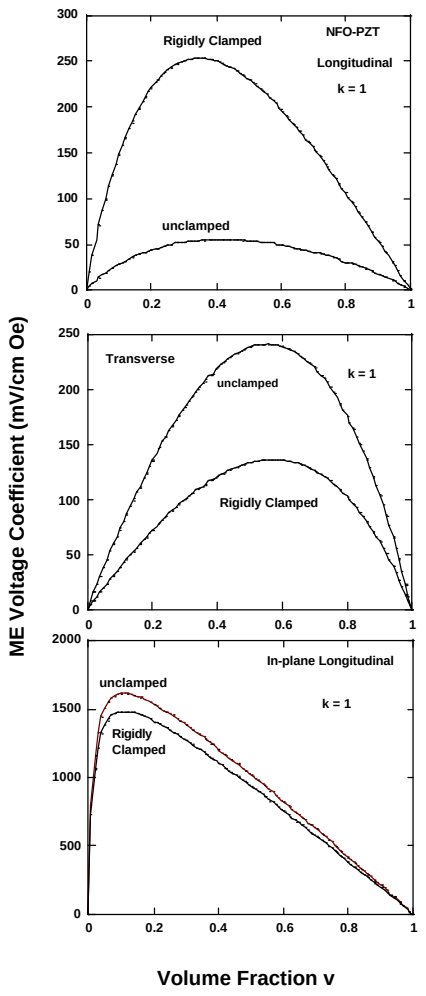

Fig. 7: Concentration *v* dependence of longitudinal, transverse and in-plane longitudinal ME voltage coefficients for unclamped and rigidly clamped nickel ferrite (NFO)-PZT bilayer for interface coupling *k*=1.

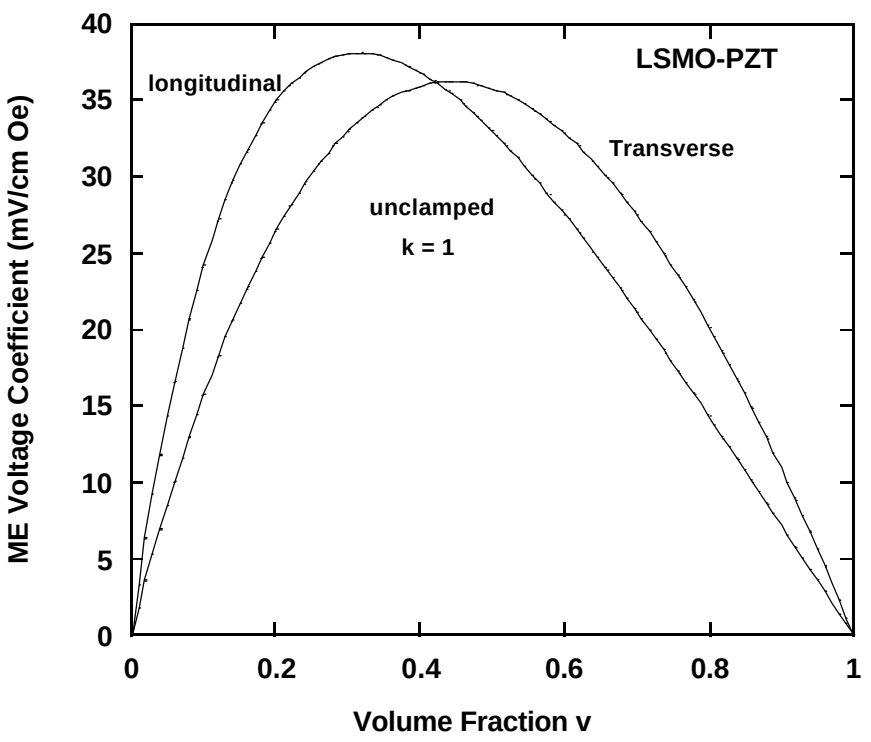

Fig. 8: Longitudinal and transverse ME voltage coefficients as a function of volume fraction *v* for PZT in unclamped $La_{0.3}Sr_{0.7}MnO_3$ (LSMO)-PZT bilayer for interface coupling *k=1.*

Finally, we consider composites with lanthanum strontium manganites for the ferromagnetic phase. Lanthanum manganites with divalent substitutions have attracted considerable interest in recent years due to double exchange mediated ferromagnetism, metallic conductivity, and giant magnetoresistance.[24] The manganites are potential candidates for ME composites because of (i) high magnetostriction and (ii) metallic conductivity that eliminates the need for a foreign electrode at the interface. We reported strong ME effects and its unique magnetic field dependence in composites of $La_{0.7}Sr_{0.3}MnO_3$ (LSMO)–PZT and $La_{0.7}Ca_{0.3}MnO_3$ (LCMO)- PZT.[8] Figure 8 shows the longitudinal and transverse ME voltage coefficients for unclamped LSMO-PZT bilayers for ideal coupling at the interface. In this case ME coefficients are quite small compared to ferrite-PZT due to weak piezomagnetic coefficients and compliances parameters for LSMO. An observation of significance is the equally strong longitudinal and transverse couplings. The magnetoelectric coupling for in-plane longitudinal fields and effects of clamping are similar in nature to ferrite-PZT bilayers and are not discussed here.

## VI. Comparison with data and discussions

It is important to compare the theoretical predictions in Figs.2-8 with data. There have been few studies on layered composites, all of them dealing with longitudinal and transverse effects on unclamped samples. To our knowledge, such measurements have never been done for in-plane longitudinal fields. Studies relevant to systems considered here include the work by Harshe, et al., on CFO-BTO and CFO-PZT,[5,12] efforts on NFO-PZT by our group[7,9] and by Lupeiko, et al.,[10] and our recent work on LSMO-PZT.[8] We also provide here our most recent data for clamped NFO-PZT.[25] Results cited here were obtained on bilayer or multilayer composites processed either by high temperature sintering or by gluing thick films/disks of ferrite/manganite-PZT. Samples were poled in an electric field and the ME coefficient was measured by subjecting the sample to a bias field H and an ac field δH. Data were obtained as a function the bias field H for a series of volume fraction *v*. Since the primary objective here is comparison of predictions of the current model with data, we refrain from any discussion on sample synthesis. Details on sample preparation and characterization are provided in Refs.5-12.

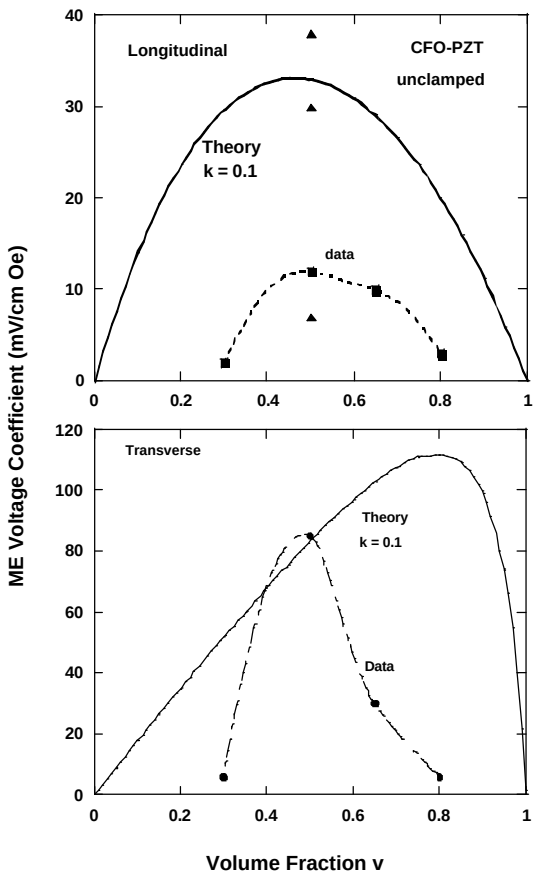

Fig. 9: Comparison of theoretical (solid curves) $a¢_E$ vs *v* and data for unclamped bilayers and multilayers of CFO-PZT. Theoretical values for *k*=0.1 for longitudinal and transverse fields are based on the present model. Data are from Ref. 12 (triangles) and Ref.9 and 17 (squares and circles).

Composites of CFO-PZT are considered first. Figure 9 shows room temperature data on low frequency (100-1000 Hz) $a¢_E$ for glued bilayers and sintered multilayers as a function of *v*. The $a¢_E$ values correspond to maximum in $a¢_E$ vs H profiles. Desired volume fractions *v* was achieved by controlling the layer thickness. The data for longitudinal and transverse fields show the anticipated increase in with *v* to a maximum, followed by a decrease for PZT-rich compositions. But the measured values in Fig.9 are an order of magnitude smaller than estimates for *k*=1in Figs.3 and 4. It is, therefore, logical to compare the data with estimated $a¢_E$ vs *v* for a reduced interface coupling *k*, on the order of 0.1. One notices general agreement between theory for *k*=0.1and data in Fig.9. The key inference here is the inherently poor interface coupling for CFO-PZT, irrespective of the sample synthesis technique. Possible causes of poor coupling are discussed later.

A similar analysis of ME interactions for NFO-PZT, however, indicated an ideal interface coupling.[7,25] Here we compare data and theory for the bias field H dependence of $a¢_{E,31}$. Figure 10 shows such data for multilayers of NFO-PZT for two conditions: unclamped and rigidly clamped samples.[7,26] As H is increased, the voltage coefficient increases to a maximum and then drops rapidly to a minimum at high fields. Since the ME coefficient is proportional to the piezomagnetic coupling, the resonance-like field dependence for $a¢_E$ is due to variations in the parameters $q_{11}$ and $q_{12}$ with H. The ME coefficient drops to zero at high fields when the magnetostriction λ attains saturation, leading to *q*=0 and the loss of ac magneto-mechanical coupling. A significant observation in Fig.10 is the *drop* in the voltage coefficient upon clamping the sample. For theoretical estimates of $a¢_E$, it is first necessary to determine the piezomagnetic coupling *q* as a function of H from magnetostriction data. Calculated values of $a¢_{E,31}$ in Fig.10 are for q-values obtained from λ vs H data in Ref.7. There is excellent agreement between theory for k=1 and data for both clamped and unclamped samples. An equally important inference in Fig.10 is the observation of a reduction in $a¢_{E,31}$ for a rigidly clamped sample, in agreement with theory (Fig.7).

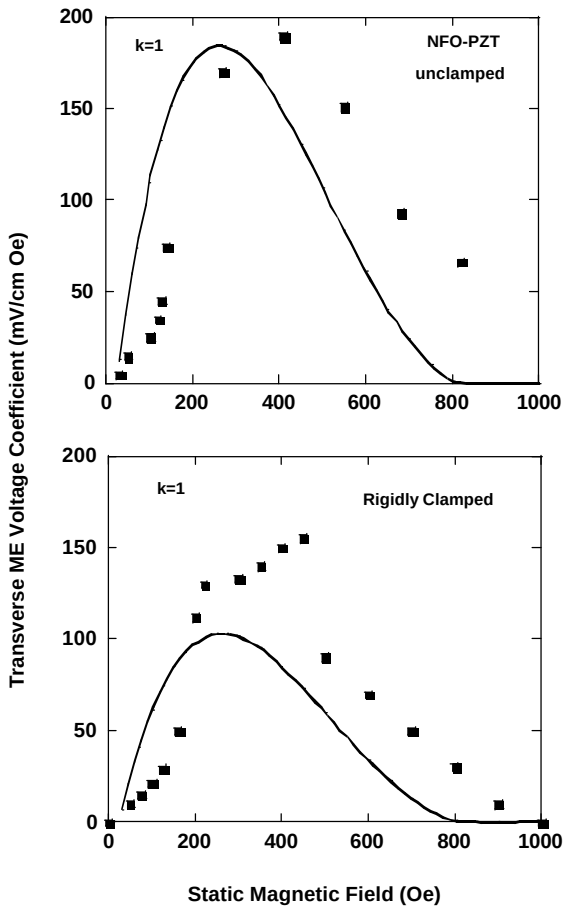

Fig. 10: Theoretical $a¢_E$ vs H and data for unclamped and rigidly clamped NFO-PZT. Theoretical estimates for *k*=1are for piezomagnetic coupling *q* obtained from magnetostriction λ vs H (Ref.7). Data (Refs.7 and 26) are for multilayer samples.

Finally for the third system of interest, LSMO-PZT, our studies revealed the weakest ME voltage coefficient among the systems considered here.[8] The ME voltage coefficient ranged from 2 to 15 mV/cm Oe, depending on the field orientations and *v*. We inferred an interface coupling k=0.2 from the analysis of $a¢_{E,31}$ vs *v* data. Thus the interface coupling is rather weak in LSMO-PZT, similar to CFO-PZT.

The comparison in Figs.9 and 10 clearly demonstrates the utility of the model discussed here for an understanding of the electromagnetic coupling in layered systems. Although the model is valid only for bilayers, data considered included both bilayers and multilayers. Now we comment on the possible cause of inferred poor coupling in CFO-PZT and LSMO-PZT and ideal coupling in NFO-PZT. The parameter *k* is expected to be sensitive to mechanical, structural, chemical and electromagnetic parameters at the interface. The high temperature sintering employed for sample processing in some studies is necessary for mechanical bonding. But the process is likely to produce microscopic structural and chemical inhomogeneities that will have adverse impact on material parameters and hence on *k*. However, it is worth noting here that a poor bonding in CFO-PZT is also inferred for glued bilayers, a process that did not involve high temperature sintering.[12] We attribute the unfavorable interface conditions to inefficient magneto-mechanical coupling. The magneto-mechanical coupling $k_m$ given by $k_m=(4\pi\lambda'\mu_r/E)^{1/2}$ where λ′ is the dynamic magnetostrictive constant and $\mu_r$ is the reversible permeability, parameters analogous to *q* and initial permeability $\mu_i$, respectively, and E is the Young's modulus.[27] In ferrites, under the influence of a bias field H and ac field δH, domain wall motion and domain rotation contribute to the Joule magnetostriction and the piezomagnetic coupling. A key requirement for strong coupling is unimpeded domain wall motion and domain rotation. A soft, high initial permeability (low anisotropy) ferrite, such as NFO, is the key ingredient for high $k_m$ and strong ME effects. Our measurements yielded an initial permeability of 20 for NFO vs 2-3 for LSMO and CFO. Thus the strong coupling in NFO-PZT is in part due to the favorable domain motion.

## VII. Conclusions

A theoretical model has been developed for low frequency ME effects in layered magnetostrictive-piezoelectric samples. Effective parameters are introduced for describing the bilayers. Expressions for the ME voltage coefficients as a function of interface coupling and volume fraction have been obtained for longitudinal, transverse and in-plane longitudinal fields. The theoretical treatment takes into consideration unclamped and clamped samples. The model has been applied to bilayers of past and current interests: CFO-BTO, CFO-PZT, NFO-PZT and LSMO-PZT. The theory predicts a giant ME effect in ferrite-PZT systems and the highest ME coupling for in-plane fields. Clamping leads to a decrease in the transverse ME coupling and an increase for longitudinal fields. Comparison with data reveal poor interface coupling for CFO-PZT and LSMO-PZT and an ideal coupling for NFO-PZT.

**Acknowledgments**

Research at Novgorod State University was supported by grants from the Russian Ministry of Education (Å02-3.4-278) and from the Universities of Russia Foundation (UNR 01.01.007). Grants from the National Science Foundation (DMR-0072144 and 0322254) supported the efforts at Oakland University.

**References**

1. L. D. Landau and E. M. Lifshitz, *Electrodynamics of Continuous Media*, Pergamon Press, Oxford (1960) p.119 (Translation of Russian Edition, 1958)
2. D. N. Astrov, Soviet phys. JETP **13**, 729 (1961).
3. J. Van Suchtelen, Philips Res. Rep., **27**, 28 (1972).
4. J. Van den Boomgaard, D. R. Terrell, and R. A. J. Born, J. Mater. Sci. **9**, 1705 (1974).
5. G. Harshe, J.O. Dougherty, and R. E. Newnham, Int. J. Appl. Electromagn. Mater. **4**, 145 (1993); M. Avellaneda and G. Harshe, J. Intell. Mater. Sys. Struc. 5, 501 (1994).
6. I.Getman, *Ferroelectrics* **162**, 393 (1994).
7. G. Srinivasan, E. T. Rasmussen, J. Gallegos, R. Srinivasan, Yu. I. Bokhan, and V. M. Laletin, Phys. Rev. B **64**, 214408 (2001).
8. G. Srinivasan, E. T. Rasmussen, B. J. Levin, and R. Hayes, Phys. Rev. B **65**, 134402 (2002).
9. G. Srinivasan, E. T. Rasmussen, and R. Hayes, Phys. Rev. B **67**, 014418 (2003).
10. T. G. Lupeiko, I. V. Lisnevskaya, M. D. Chkheidze, and B. I. Zvyagintsev, Inorganic Materials **31**, 1245 (1995).
11. J. Ryu, A. V. Carazo,K. Uchino, and H. Kim, Jpn. J. Appl. Phys. **40**, 4948 (2001).
12. G. Harshe, *Magnetoelectric effect in piezoelectric-magnetostrictive composites*, PhD thesis, The Pennsylvania State University, College Park, PA, 1991 and references therein.
13. M.I.Bichurin, Ferroelectrics, **279-280, 385** (2002).
14. M.I.Bichurin and V.M.Petrov, *Zh. Tekh. Fiz*. **58**, 2277 (1988). [Sov. Phys. Tech. Phys. **33**, 1389 (1988)].
15. M.I. Bichurin, V.M. Petrov, G. Srinivasan, *Ferroelectrics,* **280**, 165 (2002).
16. M. I. Bichurin, V. M. Petrov, and G. Srinivasan, *Bull. Am. Phys. Soc*. **47**, 772 (2002).
17. M. I. Bichurin, V. M. Petrov, and G. Srinivasan, *J. Appl. Phys.* **92**, 7681 (2002).
18. M.I. Bichurin, I. A. Kornev, V. M. Petrov, A. S. Tatarenko, Yu. V. Kiliba, and G. Srinivasan. *Phys. Rev. B* **64**, 094409 (2001).
19. M.I. Bichurin, V. M. Petrov, Yu. V. Kiliba, and G. Srinivasan. *Phys. Rev. B* **66**, 134404 (2002).
20. B. K. Sinha, W. J. Tanski, T. Lukaszek, and A. Ballato, J. Appl. Phys. **57**, 767 (1985).
21. Landolt-Bornstein; *Numerical data and functional relationships in science and technology, Group III, Crystal and Solid State Physics,* vol 4(b), *Magnetic and Other Properties of Oxides*, eds. K.-H. Hellwege and A. M. Springer, Springer-Verlag, New York (1970).
22. *Piezoelectric ceramics materials properties*, document code 13085, American Piezo Ceramics, Inc., Mackeyville, PA, (1998).
23. E. Fischer, G. Gorodetsky, and R. M. Hornreich, Solid State Comm., **10**, 1127 (1972) and references therein.
24. A. P. Remirez, J. Phys.:Condens. Matter **9**, 8171 (1997).
25. G. Srinivasan, R. Hayes and C. De Vreugd, submitted to Solid State Communications.
26. G. Srinivasan and C. De Vreugd, unpublished.
27. C. M. Van der Burgt, Philips Res. Reports **8**, 91 (1953).